\documentclass[aps,prl,twocolumn,superscriptaddress,10pt,showkeys,showpacs]{revtex4-1}
\usepackage{graphicx}

\begin{document}
\title{Strong reduction of the effective radiation length in an axially oriented scintillator crystal}

\author{L. Bandiera}
\affiliation{INFN Sezione di Ferrara $\&$ Dipartimento di Fisica e Scienze della Terra, Universit{\`a} degli Studi di Ferrara Via Saragat 1, 44122 Ferrara, Italy}
\author{V.V. Tikhomirov}
\affiliation{Institute for Nuclear Problems, Belarusian State University, Minsk, Belarus}
\author{M. Romagnoni}
\affiliation{INFN Sezione di Ferrara $\&$ Dipartimento di Fisica e Scienze della Terra, Universit{\`a} degli Studi di Ferrara Via Saragat 1, 44122 Ferrara, Italy}
\author{N. Argiolas}
\affiliation{INFN Sezione di Legnaro$\&$ Dipartimento di Fisica e Astronomia, Universit{\`a} degli Studi di Padova, Padova, Italy}
\author{E. Bagli}
\affiliation{INFN Sezione di Ferrara $\&$ Dipartimento di Fisica e Scienze della Terra, Universit{\`a} degli Studi di Ferrara Via Saragat 1, 44122 Ferrara, Italy}
\author{G. Ballerini}
\affiliation{INFN Sezione di Milano Bicocca $\&$ Dipartimento di Scienza e Alta Tecnologia, Universit{\`a} degli Studi dell'Insubria Via Valleggio, Como, Italy}
\author{A. Berra}
\affiliation{INFN Sezione di Milano Bicocca $\&$ Dipartimento di Scienza e Alta Tecnologia, Universit{\`a} degli Studi dell'Insubria Via Valleggio, Como, Italy}
\author{C. Brizzolari}
\affiliation{INFN Sezione di Milano Bicocca $\&$ Dipartimento di Scienza e Alta Tecnologia, Universit{\`a} degli Studi dell'Insubria Via Valleggio, Como, Italy}
\author{R. Camattari}
\affiliation{INFN Sezione di Ferrara $\&$ Dipartimento di Fisica e Scienze della Terra, Universit{\`a} degli Studi di Ferrara Via Saragat 1, 44122 Ferrara, Italy}
\author{D. De Salvador}
\affiliation{INFN Sezione di Legnaro$\&$ Dipartimento di Fisica e Astronomia, Universit{\`a} degli Studi di Padova, Padova, Italy}
\author{V. Haurylavets}
\affiliation{Institute for Nuclear Problems, Belarusian State University, Minsk, Belarus}
\author{V. Mascagna}
\affiliation{INFN Sezione di Milano Bicocca $\&$ Dipartimento di Scienza e Alta Tecnologia, Universit{\`a} degli Studi dell'Insubria Via Valleggio, Como, Italy}
\author{A. Mazzolari}
\affiliation{INFN Sezione di Ferrara $\&$ Dipartimento di Fisica e Scienze della Terra, Universit{\`a} degli Studi di Ferrara Via Saragat 1, 44122 Ferrara, Italy}
\author{M. Prest}
\affiliation{INFN Sezione di Milano Bicocca $\&$ Dipartimento di Scienza e Alta Tecnologia, Universit{\`a} degli Studi dell'Insubria Via Valleggio, Como, Italy}
\author{M. Soldani}
\affiliation{INFN Sezione di Milano Bicocca $\&$ Dipartimento di Scienza e Alta Tecnologia, Universit{\`a} degli Studi dell'Insubria Via Valleggio, Como, Italy}
\author{A. Sytov}
\affiliation{INFN Sezione di Ferrara $\&$ Dipartimento di Fisica e Scienze della Terra, Universit{\`a} degli Studi di Ferrara Via Saragat 1, 44122 Ferrara, Italy}
\affiliation{Institute for Nuclear Problems, Belarusian State University, Minsk, Belarus}
\author{E. Vallazza}
\affiliation{INFN Sezione di Trieste}

\date{\today}

\begin{abstract}
We measured a considerable increase of the emitted radiation by 120 GeV/c electrons in an axially oriented lead tungstate scintillator crystal, if compared to the case in which the sample was not aligned with the beam direction. This enhancement resulted from the interaction of particles with the strong crystalline electromagnetic field. The data collected at the external lines of CERN SPS were critically compared to Monte Carlo simulations based on the Baier Katkov quasiclassical method, highlighting a reduction of the scintillator radiation length by a factor of five in case of beam alignment with the [001] crystal axes. The observed effect opens the way to the realization of compact electromagnetic calorimeters/detectors based on oriented scintillator crystals in which the amount of material can be strongly reduced with respect to the state of the art. These devices could have relevant applications in fixed-target experiments as well as in satellite-borne $\gamma$-telescopes.
\end{abstract}


\maketitle

Since their discovery, scintillator materials have played an important role in nuclear and particle physics, as well as in medical and industrial imaging. In particular, inorganic scintillator crystals are widely exploited for the realization of homogeneous electromagnetic (e.m.) calorimeters for high-energy physics (HEP) and astrophysics \cite{CMS, Atw}, to measure the energy of $e^{\pm}$ and of $\gamma$-rays. The process of energy measurement via e.m. calorimeters is related to the e.m. shower development with generation of secondary particles, consisting in a long chain of events of $\gamma$ emission by $e^\pm$ and $e^+ e^-$ pair production (PP) by $\gamma$. Scintillation light produced by the passage of particles is proportional to energy deposited inside the crystal and is collected by photo-detectors. In order to measure the initial particle energy, the whole shower should be contained in the detector. Since for primary particles of multi-GeV or TeV energies the shower results to be ten or more radiation lengths ($X_0$) long, high-Z scintillator crystals (e.g., BGO ($Bi_4Ge_3O_{12}$), CsI, and PWO ($PbWO_4$)) with $X_0$ of about 1 cm have been introduced to realize compact calorimeters.

Despite these materials are crystalline, the lattice influence on the e.m. shower is usually completely ignored both in detector design and simulations. On the other hand, it is well known since 1950s that the crystal lattice may strongly modify both $\gamma$ emission by $e^{\pm}$ and PP by $\gamma$, thus accelerating the e.m. shower development. Firstly, the interference effects in bremsstrahlung and PP accompanying $e^{\pm}$ and $\gamma$ propagation at relatively small angles with respect to crystal planes/axes---the so-called coherent bremsstrahlung (CB) and coherent PP (CPP)---were investigated \cite{Fer,TM,Uber}. CB consists in the enhancement of bremsstrahlung when the momentum transferred by $e^\pm$ to the crystal matches a reciprocal lattice vector, in analogy with Bragg/Laue diffraction, and can be described as a first order perturbation of the particle motion within the straight trajectory approximation. But, when the particle velocity is nearly parallel to either a crystal axis or a plane, the straight trajectory approximation is no longer valid, since in this case a charged particle would be forced in an oscillatory motion within the planar/axial potential well, i.e. the \textit{channeling} phenomenon, leading to a specific e.m. radiation emission, thereby called \textit{channeling radiation}.
Indeed, for small angle between charged particle trajectory and crystal axes/planes direction, successive collisions of the particles with the atoms in the same plane/row are correlated and it is possible to replace the screened Coulomb potential of each atom with an average continuous potential of the whole plane/string \cite{Lin}, corresponding to a strong electric field $E \sim 10^{10}-10^{12}$ V/cm \cite{Bar,Bai}. Channeling may occur if the incidence angle of the charged particle with respect to the crystal planes/axes, $\psi$, is lower than the critical value introduced by Lindhard $\theta_c = \sqrt{2V_0/\varepsilon}$, where $\varepsilon$ is the particle energy and $V_0$ the average potential energy amplitude. The continuous potential can be used also to describe dynamics and radiation processes for unchanneled particles that enter the crystal with $\psi \ge \theta_c$. Indeed, such particles are subject to an oscillatory motion while crossing successive planes/axes with a period equal to $d/\psi$, $d$ being the interplanar/interaxial distance.

For sub-GeV particles, the channeling radiation process is undulator-like with emission of relatively soft photons, while at few-GeV energies the $e^\pm$ deflection angle in the oscillatory motion starts to exceed the typical radiation cone opening angle, $1/\gamma$, where $\gamma$ is the Lorentz factor ($\gamma=\varepsilon/m$) and $m$ the $e^\pm$ rest mass. As a consequence, the radiation process switches from dipole (typical also for bremsstrahlung and CB) to synchrotron regime, for which both radiation intensity and emitted photon energy increase proportionally to $\varepsilon^{2}$ \cite{Bar,Kim,Bai,Sor,Ugg,LL4,Bai}. The synchrotron regime holds true also for unchanneled particle with incidence angle $\psi < V_0/m$ \cite{Bai}.

At even higher energy, the average electric field felt by a particle in its rest frame is enhanced by a factor $\gamma$ because of the Lorentz contraction, thus becoming comparable to the Schwinger critical field of QED, $E_0 = m^2 c^3/e\hbar = 1.32\cdot 10^{16} V/cm$, which can be reached for instance in pulsar atmosphere. This \textit{strong field} regime is characterized by quantum synchrotron radiation with intense hard photon emission by $e^{\pm}$, as well as by intense PP by high-energy photons \cite{Bar}. The \textit{strong field} limit is attainable in crystals if the $e^{\pm}$/$\gamma$ beam energy reaches tens/hundreds of GeV, when the parameter $\chi = \gamma E/E_0$ is $\sim 1$.

The huge enhancement of radiation and PP processes caused by the \textit{strong crystalline field} was predicted by the authors of \cite{Bai,Bar,Kim,Sor} and observed at CERN and IHEP since middle 80s with the usage of single crystals of diamond, Si, Ge, W, and Ir \cite{Bel,Bel2,Med,Bas,Bau}. The main consequence of these effects was the acceleration of the e.m. shower development, which resulted in a strong reduction of the shower length and thereby of the radiation length, $X_0$, in comparison with amorphous materials (or equivalently randomly-oriented crystals).

Since high-Z crystal scintillators are widely exploited in e.m. calorimeters for HEP, it became rather important to investigate the \textit{strong field} influence on the e.m. shower development in these materials. The only previous study in this direction was carried out about 20 years ago with oriented garnet and PWO crystals at intermediate electron energy (26 GeV, corresponding to $\chi ~\simeq 1$), which resulted in a limited increase in energy loss of about 10\% of the beam energy as compared to random orientation \cite{Bas2}. The authors claimed the importance to further investigate such processes at higher energies (about hundreds-GeV or TeV), where the strong field effects would become more important and at which current and future HEP experiments would work. The need of a full Monte Carlo code to simulate the e.m. shower development in oriented crystals was also highlighted.

In this letter we present an experimental investigation of the e.m. radiation generated by a 120 GeV/c electron beam under axial alignment with a PWO crystal, demonstrating the possibility to strongly reduce the effective radiation length as compared to the unaligned case. This effect could be advantageously exploited to reduce the amount of scintillator material in future e.m. calorimeters and $\gamma$-telescopes, as discussed later in the text.
	
\begin{figure}
\begin{center}
\includegraphics[width=\columnwidth]{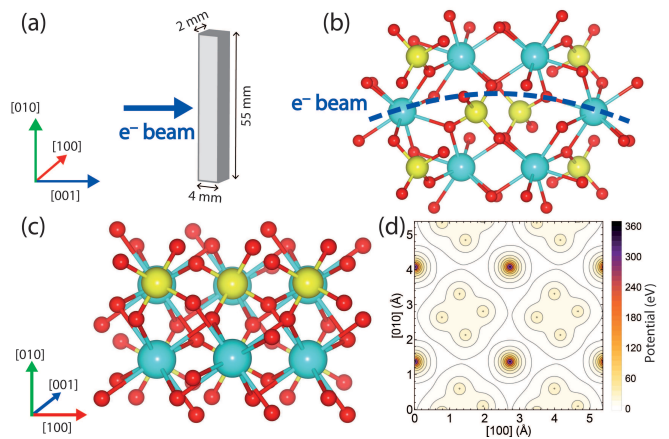}
\caption{PWO crystal and its electric potential. (a) Orientation of the crystal with respect to the electron beam. PWO crystal lattice from the side (b) and along the trajectory (c). (d) Average continuous axial potential felt by electrons moving along the [001] axis. Room-temperature thermal vibrations have been taken into account.}\label{fig:PWO}
\end{center}
\end{figure}

A PWO crystal has a scheelite-type structure characterized by a tetragonal lattice with constants \emph{a} = \emph{b} = 5.456 $\textup{\AA}$ and \emph{c} = 12.020 $\textup{\AA}$. The single crystal used for this experiment was a strip-like sample 2$\times$55$\times$4$ $mm$^3$ large, with the largest faces oriented parallel to the (100) planes. The electron beam was aligned with the [001] crystal axis, which was parallel to the $L =$ 4 mm $=$ 0.45$X_0$ crystal side, as shown in Fig.\ref{fig:PWO}(a). Fig.\ref{fig:PWO} also displays the PWO crystal lattice from [100] (b) and [001] (c) axis view, while the continuous axial potential felt by the particle moving along the [001] axes is represented in Fig.\ref{fig:PWO}(d). The high-energy electron beam of the CERN SPS H4 beamline was used to explore the deep strong field regime, characterized by a factor $\chi \simeq 4$, corresponding to a maximal axial field strength of $E \approx 2.3\times 10^{11}V/cm$ and a Lorentz factor $\gamma \approx 2.35 \times 10^{5}$. The crystal thickness was chosen to highlight the transition from the "thin target" limit ($L < X_0$) for the case of random crystal-to-beam alignment, to the opposite condition of "thick target", corresponding to the crystal disposed in axial orientation, in which $X_0$ is reduced to an effective value, $X_{0_{eff}}$, in such a way that $L > X_{0_{eff}}$.

\begin{figure}
\includegraphics[width=\columnwidth]{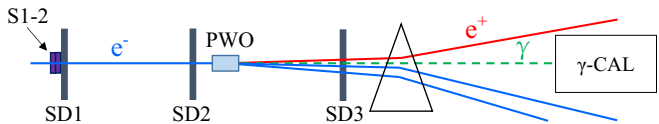}
\caption{Experimental setup. S1-2 are the plastic scintillators used for the trigger; SD1, SD2, and SD3 are the Silicon Detectors forming the tracker. The triangle represents the bending magnet that separates both the primary electrons and secondary charged particles produced by emitted photons inside the PWO sample. $\gamma$-CAL is the e.m. calorimeter used to collect the emitted photons.}\label{fig:setup}
\end{figure}

A sketch of the setup is presented in Fig.\ref{fig:setup}. The setup described in \cite{Ban} was upgraded with a new $\gamma$-calorimeter based on a $3\times 3$ matrix of CMS-type ECAL crystals, covering a total of 25$X_0$ to measure all the produced $\gamma$-rays downstream the crystal. Also the $\gamma$-CAL, in turn, was made of lead tungstate. The PWO sample was mounted on a high-precision goniometer with a resolution of a few $\mu rad$ in both horizontal and vertical axial rotations. The trajectories of the electrons interacting with the crystal have been reconstructed through a telescope system based on high precision micro strip detectors ($SD1$, $SD2$, and $SD3$ in Fig.\ref{fig:setup}) with an angular resolution of about 10 $\mu rad$. The beam divergence was 80$\times$90 $\mu$$rad^2$ in horizontal and vertical directions, respectively. After the silicon telescope, a bending magnet (the triangle in Fig.\ref{fig:setup}) separated the charged beam from the emitted $\gamma$-rays, which are collected at the downstream e.m. calorimeter.

\begin{figure}
\includegraphics[width=\columnwidth]{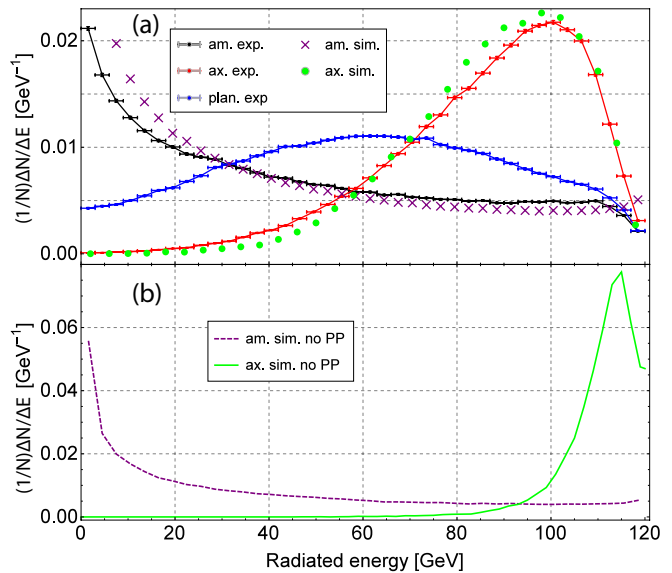}
\caption{a. Distributions of the radiated energy by 120 GeV electrons in the 4 mm-long PWO target, $(1/N)\Delta N(E)/\Delta E$, measured collecting all the survived photons with the $\gamma$-CAL placed downstream the crystal, in three different crystal-to-beam configuration: random (am. exp.), axial (ax. exp.), and planar (pl. exp.) alignment. The Monte Carlo simulations regarding the axial and amorphous case (ax. sim. and am. sim.) including in the code both the PP by the emitted $\gamma$ and $\gamma$ emission by the produced $e^\pm$ are also shown. b. Simulated total radiative energy loss obtained by switching off the PP by the emitted $\gamma$ for the randomly (am. sim. no PP) and axially (ax. sim. no PP) aligned cases.}\label{fig:energy}
\end{figure}

We firstly aligned the crystal with respect to the $(100)$ planes by exploiting the horizontal rotational movement of the goniometer. Then, the crystal was aligned with the [001] axes by scanning the vertical rotational movement. In particular, the PWO sample was placed in three different configurations, i.e. with the beam entering the crystal disposed in axial alignment (ax. exp.), in planar alignment (pl. exp.) and in random alignment, which is equivalent to the amorphous case (am. exp.). Fig.\ref{fig:energy}-a displays the experimental radiated energy distribution of the 120 GeV/c electrons $(1/N)\Delta N(E)/\Delta E$, where $\Delta N(E)$ is the number of events acquired in the range $[E - \Delta E/2, E + \Delta E/2]$, $\Delta E$ being the bin size of the distribution and $N$ the total number of entries. The x-axis values represent the sum of the energies of all the photons collected at the downstream $\gamma$-CAL; in other words, all the produced $\gamma$s by each primary $e^{-}$ in the PWO crystal and not absorbed/converted inside it. Indeed, the crystal was thick enough to allow pairs production from emitted photons and, since it is not possible to distinguish the primary from the secondary electrons, all of them were swiped away by the magnetic field upstream the $\gamma$-CAL (see Fig. \ref{fig:setup}). Details of the measurement procedure can be found in Ref. \cite{Lie,Ban2}. As expected, the distributions for both the axial and planar cases are harder if compared with the random case and peaked at 60 and 100 GeV, respectively. On the other hand, the radiated energy distribution related to the amorphous case has the typical shape of nearly single-photon Bethe-Heitler spectrum, with a sharp $1/E$-like decrease in the soft photon region, on the left.

The enhancement of $\gamma$ quanta emission resulted in a decrease of the effective radiation length in the axially oriented PWO crystal, with an acceleration of the e.m. shower development. Further proof of this is the relevant growth of production of secondary charged particles, as measured with the SD3 silicon detector placed downstream the crystal just before the bending magnet. During the analysis, we selected only single hits registered by both SD1 and SD2 detectors, i.e., only incoming single tracks onto the crystal. It resulted that the number of produced particles detected by the Si detector SD3 was enlarged by a factor of three when the sample was axially aligned, with only 30\% left of single tracks if compared with the 70\% in case of random orientation. This was indeed a first qualitative demonstration of the transition from nearly pure bremsstrahlung to pronounced shower development, with an increase of generation of secondary particles, caused by the strong decrease of $X_{0_{eff}}$.

Since part of the radiated energy by the primary electrons was converted in pairs and not collected at the calorimeter, a proper estimation of the $X_{0_{eff}}$ reduction factor can be done only through Monte Carlo simulation. We recently developed a method of direct integration of the Baier Katkov (BK) formula \cite{Gui,Ban} and tested it vs. experimental results with Si crystals \cite{Ban,Ban2,Ban3}. The BK quasiclassical method is used to compute the e.m. radiation emission by ultrarelativistic particles in an external field taking into account quantum recoil and is commonly used to treat the strong field effects in crystals \cite{Artru}. For the case of this letter, we included the contribution of PP through the BK method to describe the shower development. The results of the simulation for the radiated energy distributions for 120 GeV $e^{-}$ under axial (ax. sim.) and random (am. sim.) alignment are shown in Fig.\ref{fig:energy}-a, being in quite good agreement with the experimental results. Indeed, with the aim of reproducing the experimental configuration, the simulation plots include only the total energy of the photons survived after the crystal.

The agreement between the experiment and simulations proved the feasibility of the method that can be used to infer the total radiative energy loss by the primary electron. This was done modifying the code by switching off the PP by the emitted $\gamma$-rays. The simulation result is shown in Fig.\ref{fig:energy}-b. Here, in case of axis-to-beam alignment the distribution shows a peak around 115 GeV, with an average energy loss $E_{loss}=109.7$ GeV per $e^{-}$. This latter value is consistent with $X_{0_{eff}} \simeq$ 1.6 $mm$, determined by the equation $E_{loss} = E_{beam}(1-e^{-L/X_{0_{eff}}})$ with $E_{beam}=120$ GeV, which results in a ratio $X_0/X_{0_{eff}} \simeq 5$. Such estimation is an approximation since the primary beam energy, and consequently the $\chi$ parameter, decrease during the shower development, with a consequent increase of $X_{0_{eff}}$ \cite{Bar2}. Nevertheless, through simulation it is possible to extrapolate the correct value at different particle energy, being indeed $X_{0_{eff}}\simeq 1.35$ $mm$ for the primary 120 GeV $e^{-}$.

\begin{figure}
\includegraphics[width=\columnwidth]{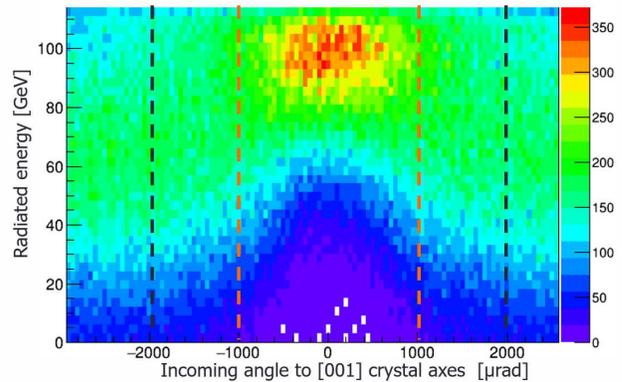}
\caption{Experimental radiated energy distribution by 120 GeV/c $e^{-}$ inside the crystal vs. [001] axes-to-beam orientation, as measured by the downstream calorimeter collected by rotating the crystal with the goniometer along the (100) planes. The region within $\pm$1 $mrad$ is between two red-dashed vertical lines, while the region within $\psi = \pm$2 $mrad$ mrad is between two black-dashed vertical lines.}\label{fig:experimental}
\end{figure}

For applications in both HEP and astrophysics, the angular acceptance of e.m. shower enhancement is a key parameter. Fig.\ref{fig:experimental} shows the experimental radiated energy distribution vs. [001] axes-to-beam orientation, collected by rotating the crystal along the (100) planes with the goniometer. The relevant increase in the energy loss due to the interaction with the crystal axes is maintained for about $\pm$ 1 $mrad$ (vertical red-dashed lines), which corresponds to the theoretical angular region for strong synchrotron-like radiation and PP, i.e $\psi \simeq \pm V_0/m = \pm 1$ $mrad$. As expected, such angular acceptance is one order of magnitude larger than the critical angle for channeling $\theta_c \approx 0.1$ $mrad$. At larger incidence angle, the distribution approaches to the planar level peaked at 60 GeV when $\psi \simeq $ 2-3 $mrad > V_0/m$. From theory and previous experiments with W crystals \cite{Moore}, at still larger incidence angle with respect to either axial or planar directions, CB/CPP regime holds true \cite{TM} and a measurable contribution to radiation/PP increase should manifest up to $\psi \sim 1^{\circ}$.

The measured $X_0$ reduction manifests in an acceleration of the e.m. shower development, the strength of which depends on the particle energy and its direction with respect to the main crystal axes, being maximal at the beginning of the shower when the primary particle possesses the highest energy \cite{Bar2}. The existing e.m. calorimeters made of high-Z scintillators, such as ECAL for CMS and Fermi LAT, have a crystalline structure \cite{CMS,Atw}. Thus, in principle, orientational effects could have been registered also for these experiments, even if the crystalline structure of the scintillators was not considered when they where mounted. However, this possibility is very unlikely because the majority of the events occurs at $\psi \gg V_0/m$, thereby making the influence of the crystal structure quite marginal. Nevertheless, the unavoidable existence of rare events with $e^\pm$ and $\gamma$ incidence at $\psi \leq V_0/m$ with one of the main crystal axes would demand further investigation.

On the other hand, the acceleration of e.m. shower development in oriented scintillator crystals can be exploited for future fixed-target experiments to build compact forward e.m. calorimeters/preshowers, with a considerable reduction of the detector thickness for beam incidence in the range $\psi \sim mrad$. Furthermore, these effects can be applied to diminish the thickness of a photon converter, taking advantage of the reduced ratio between $X_0$ and nuclear interaction length, thus increasing the transparency to hadrons as done in the past by the NA48 experiment at CERN \cite{biino}, as well as in the construction of a downstream calorimeter to detect photons in the existing beam without being blinded by hadronic interactions \cite{klever}. A similar principle can be transferred to light dark matter search to decrease the dump length, which is a crucial parameter \cite{raggi}. Indeed, if a dark photon is created during the e.m. shower initiated by a primary $e^{\pm}$, it can be detected only if survives after the remaining dump length. The shorter is such length, the higher is the sensitivity to dark photons.

Furthermore, with the birth of multimessenger astrophysics, one may think of pointing a telescope towards a source, thus measuring the spectrum of a point-like $\gamma$-source in the TeV energy region \cite{Bai,UggE}. If a satellite is equipped with a calorimeter module made of oriented crystals, the shower of $\gamma$-rays with energy larger than 100 GeV can be completely contained in a quite compact volume, reducing the necessary weight---and therefore the cost---compared to those currently used. In the absence of pointing, a calorimeter made of oriented scintillators would continue to operate in a standard way. Thinking of an apparatus similar to FERMI LAT \cite{Atw}, the reduction of $X_0$ in oriented crystals could be useful to reduce both the calorimeter volume and the thickness of the photo-converters in the tracking system, thus reducing the dispersive effect of the multiple scattering that worsens the detector resolution. The advantage of using high-Z scintillators instead of metallic crystals, such as W or Ir as NA48 did, is the better crystallographic quality and the possibility to be grown in virtually any size \cite{Bas2}. Indeed, the mosaic spread of the PWO sample used in the experiment is $\approx$ 0.1 $mrad$ as measured with X-ray diffraction, while it is usually more than 1 mrad for commercially available metallic crystals. Moreover, the transparency of the crystals gives the additional possibility to measure directly the characteristics of the cascade by collecting the scintillation light, as in the CERN NA64 experiment \textit{active beam dump} \cite{NA64}.

In summary, the experimental evidence of strong enhancement of the e.m. radiation generated by 120 GeV/c electrons in an axially oriented lead tungstate scintillator crystal, with respect to the case in which the sample was not aligned with the beam direction, has been demonstrated at the CERN SPS external lines. A relevant increase of secondary particles production inside the crystal also proved the acceleration of the e.m. shower development, leading to a decrease of the effective radiation length by several times as reproduced by Monte Carlo simulations. Future studies are required to characterize the shower development of hundreds-GeV primary particles by directly collecting the scintillation light vs. the crystal-to-beam orientation. The radiation length reduction in axially oriented scintillator crystals can have important repercussions in the realization of forward calorimeters/preshowers or $\gamma$-converters with a limited thickness to be exploited in fixed-target experiments in high-energy physics. Furthermore, one may also take advantage of this effect to realize compact satellite-borne $\gamma$-telescopes for pointing strategy in the GeV/TeV energy range.

This work was partially supported by the INFN-AXIAL experiment and by the EU commision through the PEARL (G.A. 690991) and CRYSBEAM (G.A. 615089) projects. This work is also partially supported by the Belarusian Republican Foundation for Fundamental Research and the Ministry of Education of the Republic of Belarus under Contract No. F18MV-026. We acknowledge the support of the PS/SPS physics coordinator and of the CERN SPS, Secondary Beams and Areas staff. We acknowledge the CINECA award under the ISCRA initiative for the availability of high performance computing resources and support. We also acknowledge technical support of A. Gaiardo and M. Valt. L. Bandiera would like to thank S. Cutini, G. Mezzadri, M. Moulson, P. Picozza, M. Raggi, and M. Tavani for fruitful discussion on possible applications and V. Guidi and G. Cavoto for funding and mentoring support.


%

\end{document}